\input{aipcheck}

\documentclass[
  ,draft            % use draft while you are working on the paper
  ]
  {aipproc}

\newcommand{\ba}{\begin{eqnarray}}
\newcommand{\ea}{\end{eqnarray}}
\newcommand{\baa}{\begin{eqnarray*}}
\newcommand{\eaa}{\end{eqnarray*}}
\newcommand{\bea}{\begin{eqnarray*}}
\newcommand{\eea}{\end{eqnarray*}}

\newcommand{\bb}{\begin{thebibliography}}
\newcommand{\eb}{\end{thebibliography}}

\newcommand{\bit}[1]{\bibitem{#1}}

\newcommand{\AmS}{{\protect\the\textfont2
  A\kern-.1667em\lower.5ex\hbox{M}\kern-.125emS}}

\newcommand{\bi}{\scriptstyle b_{1}}

\newcommand{\hi}{\scriptscriptstyle h_{1}}

\newcommand{\no}{\nonumber}

\newcommand{\nn}{\nonumber \\}

\newcommand{\delv}{\delta q^v}

\newcommand{\ai}{\scriptstyle a_{1}}

\newcommand{\bgi}{\begin{itemize}}
\newcommand{\eni}{\end{itemize}}

\newcommand{\be}{\begin{eqnarray}}
\newcommand{\ee}{\end{eqnarray}}

\newcommand{\bslash}{{\cal B} \hskip -0.6em /}

\newfont{\fib}{cmfi10 at 10pt}

\newcommand{\eg}{{\it e.g.}\ }

\layoutstyle{6x9}

\begin{document}

\title{Transversity in Exclusive and Inclusive Processes}

\author{Leonard Gamberg}{
  address={Division of Science,
Penn State-Berks Lehigh Valley College, Reading, PA 19610,
USA}
\footnote{
Supported by a Research Development
Grant, Penn State Berks. 
Based on talk given at $15^{\rm th}$
International Spin Physics Symposium (SPIN 2002), 9-14 September, 2002.}}

\author{Gary R. Goldstein}{
  address={Department of Physics
          and Astronomy, Tufts University,
           Medford, MA 02155, USA}
\footnote{Supported by the US Department
of Energy DE-FG02-29ER40702.}
}

\author{Karo A. Oganessyan}{
 address={INFN-Laboratori Nazionali di Frascati, Enrico Fermi 40,
  I-00044 Frascati, Italy}
  ,altaddress={DESY, Notkestrasse 85, 22603 Hamburg, Germany}
}

\begin{abstract}
Both meson photoproduction and semi-inclusive deep
inelastic scattering can potentially probe transversity
properties of the nucleon.  We explore how that potential can be
realized dynamically.  The role of rescattering in
both exclusive and inclusive meson production as a source
for single spin asymmetries is examined.
Using a dynamical model, we evaluate the spin independent $\cos 2\phi$
asymmetry associated with transversity of quarks inside unpolarized
hadrons, at HERMES kinematics.
We also explore the effects of rescattering on the 
transversity distribution of the nucleon.
\end{abstract}

\maketitle

%%%%%%%%%%%%%%%%%%%%%%%%%%%%%%%%%%%%%%%%%%%%
%% MAINMATTER
%%%%%%%%%%%%%%%%%%%%%%%%%%%%%%%%%%%%%%%%%%%%

\subsection{Transversity}
\vskip-0.25cm
It is  well known that 
the  leading twist transversity distribution $h_1(x)$~\cite{jaffe91} 
and its first moment, the tensor charge, being  chiral
odd   cannot be accessed in deep inelastic scattering.
However, $h_1(x)$ can be probed when  at least two hadrons are
present, \eg Drell Yan~\cite{ralston79} or
semi-inclusive deep inelastic scattering (SIDIS). In the
latter process at leading twist, the property of quark transversity
can be measured via the azimuthal asymmetry
in the fragmenting hadron's momentum and spin distributions.
For example, spinless hadrons produced in
the so-called Collins asymmetry~\cite{cnpb92}
depend on the transverse momentum of quarks in the target, $k_T$,  
and fragmentation functions, $p_T$~\cite{kotz0}.  
Including transverse momentum
leads to an increase in the number of
leading twist distribution and fragmentation 
functions (\eg $h_{1T}(x,k_T), h_{1T}^\perp(x,k_T), H^{\perp}_1(z,p_T)$).
Allowing time reversal odd ($T$-odd) quark distribution 
functions~\cite{ANSL,cplb,mulders2},  
$h_1^\perp(x,k_T), \ f_{1T}^\perp(x,k_T)$
suggests they  enter the semi-inclusive  unpolarized momentum,
and polarized spin, asymmetries. 
That is, the distribution $f_{1T}^\perp(x,k_T)$ representing 
the number density of {\em unpolarized} quarks in transversely polarized
nucleons, maybe  entering the recent measurements of SSAs at
HERMES and SMC in semi-inclusive pion electroproduction~\cite{hermes}.
Alternatively, $h_1^\perp(x,k_T)$ which 
describes the transfer of  transversity to quarks inside unpolarized
hadrons may enter transverse momentum dependent asymmetries.
Beyond the $T$-odd properties, the existence of these distributions
are a signal of the {\em essential} role 
played by the intrinsic transverse quark
momentum and the corresponding angular momentum of quarks inside
the target and fragmenting hadrons  in these hard scattering processes.

Further insight into transversity has come from analyzing 
quark-target helicity flip 
amplitudes in deeply virtual Compton
scattering (DVCS)\cite{diehl}.  
%Written as 
%combinations of the generalized parton distribution, $H(x,\xi,t)$
%reduce in the forward limit to $h_1(x)$.  
Angular momentum conservation 
requires that helicity changes are
accompanied by transferring 1 or 2 units of 
orbital angular momentum; {\em again highlighting}
the essential role of intrinsic $k_{T}$ and
orbital angular momentum in determining transversity. 

\subsection{The Exchange Picture}
The interdependence of transversity on
quark {\em orbital} angular momentum and $k_{T}$
is more general than suggested in the above discussion 
on SSAs and the GPD analysis of
transversity. This behavior arises
in ref.~\cite{gamb_gold} where we study the
vertex function associated with the tensor charge
in exclusive meson production.
Again, angular momentum conservation results in  the transfer of
orbital angular momentum $\ell=1$ carried by the dominant
$J^{PC}=1^{+-}$
mesons  to compensate for the non-conservation of helicity across
the vertex.  Transverse momentum dependence arises from
the axial vector mesons that {\em dominate} the
tensor coupling (They are the $C$-odd -- $h_1$(1170), $h_1$(1380),
$b_1$(1235)).
These mesons are in the $\left(35\otimes \ell=1\right)$
multiplet of the
$SU(6)\otimes O(3)$ symmetry group that best represents the
mass symmetry among the low lying mesons.
Along with axial vector
dominance this symmetry results in the isoscalar and isovector
contribution to the tensor charge
{\small\be
\delta u(\mu^2) -\delta d(\mu^2)
=\frac{5}{6}\frac{g_A}{g_V}\frac{ M_{\ai}^2}{
M_{\bi}^2}\frac{\langle k_{T}^2\rangle}{ M_N M_{\bi}},
\quad \delta u(\mu^2)+\delta d(\mu^2)
=\frac{3}{5}\frac{M_{\bi}^2}{M_{\hi}^2}\delv .
\ee}
Each depends on two powers of the average intrinsic quark
momentum $\langle k_{T}^2\rangle$, because  the tensor couplings
involve helicty flips associated with  kinematic factors
of $3$-momentum transfer.
as required by angular momentum conservation.

The $k_{T}$ dependence can be understood on fairly general grounds
from the kinematics of the exchange picture
in exclusive pseudoscalar meson photoproduction. For large
$s$ and relatively small momentum transfer $t$ simple combinations of
the four helicity amplitudes involve definite parity exchanges.
The four independent helicity amplitudes can have
the minimum kinematically allowed powers,
{\small\be
f_1 = f_{1+,0+} \propto k_{T}^1 , \quad
f_2 = f_{1+,0-} \propto k_{T}^0 ,\quad
f_3 = f_{1-,0+} \propto k_{T}^2 , \quad
f_4 = f_{1-,0-} \propto k_{T}^1.  \nonumber
\ee}
However, in single hadron exchange (or Regge pole exchange)
parity conservation requires {\small $f_1 =\pm f_4 \quad {\rm and} \quad f_2 =\mp f_3$} for even/odd parity exchanges. These pair relations,
along with a single hadron exchange model,
force $f_2$ to behave like  $f_3$ for small
$t$. This introduces the $k_{T}^2$ factor into $f_2$. However for a
non-zero polarized target asymmetry  to arise there must be
interference between single helicity flip and non-flip and/or double
flip amplitudes. Thus this asymmetry must arise from rescattering
corrections (or Regge cuts or eikonalization or loop corrections) 
to single hadron exchanges. That is, one of the amplitudes in
{\small\be
P_y = \frac{2 Im (f_1^*f_3 - f_4^*f_2)}
{\sum_{j=1...4}|f_j|^2}
\ee}
must acquire a different phase.
In fact rescattering  reinstates $f_2 \propto k_{T}^0$ 
by integrating over loop $k_{T}$, which
effectively introduces a $\left< k_{T}^2 \right>$
factor~\cite{gold_owens}. This is true for the {\em inclusive process} 
as well, where only one final hadron is measured; a relative phase
in a helicity flip three body amplitude is required.
\subsection{Rescattering and SIDIS}
\vskip-.25cm
Recently a rescattering approach was applied to the calculation of
SSA in pion electroproduction, 
using a QCD motivated quark-diquark model of the 
nucleon~\cite{brodsky} (BHS).
In Ref.~\cite{ji,cplb} the rescattering effect
is interpreted as giving rise to the $T$-odd 
Sivers~\cite{sivers}  $f_{1T}^\perp$
function; the number density of unpolarized quarks in a transversely 
polarized target. 
Being $T$-odd, this asymmetry vanishes at
tree level. The important lesson beyond the model calculation, is that,
theoretically, final state interactions are essential for producing
non-zero SSAs. Furthermore, the phenomenological determination of quark
spin distributions can be disentangled from measurements of SSAs.

We have investigated the 
rescattering contributions to the transversity distribution $h_1(x)$ and
the $T$-odd function $h_1^\perp(x)$ 
and corresponding asymmetries in SIDIS.  Collins~\cite{cnpb92} 
considered one such process, the production of pions from 
transversely polarized quarks in a transversely polarized target.
The corresponding SSA involves the convolution of
the transversity distribution function and the $T$-odd fragmentation
function, $h_1(x)\star H_1^\perp(z)$~\cite{kotz,boer}. The 
transversity distribution function is  defined through the light-cone
quark distribution with gauge link indicated,
{\small\be
s_T^i\Delta f_T(x,k_{T})&=&{\frac{1} {2}}\sum_n
\int {\frac{d\xi^- d^2\xi_\perp }
  {(2\pi)^3}} e^{-i(\xi^- k^+-\vec{\xi}_\perp \vec{k}_\perp)} 
\langle P|\overline{\psi}(\xi^-,\xi_\perp)|n\rangle
\nn &&
\langle n|\left(-ie_1\int^\infty_0 A^+(\xi^-,0) d\xi^-
   \right)\gamma^+ 
\gamma^i\gamma^5\psi(0)|P\rangle  + {\rm h. c.},
\ee}
where $e_1$ is the charge of the struck quark and
$n$ represents intermediate diquark states. 
Integration over $q^\mu$, the gluon momentum 
we obtain
{\small\be
&&s_{T}^i\Delta
f_T(x,k_{T})=\frac{e_1e_2g^2}{2(2\pi)^4}
\frac{1-x}{\Lambda(k_T^2)}
\left\{\left(S^i_T\Big[\bigg(m+xM\bigg)^2+k_T^2\bigg]
+2k^i_{T}\mathbf{S}_T\cdot\mathbf{k}_T\right)\right.
\no \\
&&\times\frac{1}{k^2_T+\Lambda(0)^2
+\lambda_g^2}
\left(\ln\frac{\Lambda(k^2_T)}{\Lambda(0)}
+\ln\frac{k^2_T+\lambda^2_g}{\lambda_g^2}\right)
-\left.
\left(S^i_T k_T^2+2k_T^i\mathbf{S}_T\cdot\mathbf{k}_T
\right)
\frac{1}{k_T^2}\ln\frac{\Lambda(k^2_T)}
{\Lambda(0)}
\right\},
\ee}
where {\small $\Lambda(k_{T}^2) = k_{T}^2 +
x(1-x)\left(-M^2+\frac{m^2}{x} + \frac{\lambda^2}{1-x}\right).$}
The (Abelian) gluon mass (usually chosen at $\lambda_g\approx 1\ GeV$)
is indicative of $\chi SB$ scale and appears here to regulate the IR
divergence.  This one loop contribution constitutes the next order term in an 
eikonalization. The first part has the same nucleon spin dependent 
structure as a tree level model calculation (modified 
by the log terms) - it  is leading twist and a 
combination of $h_{1T}(x,k_{T})$ and $h_{1T}^{\perp}(x,k_{T})$. 
The second part has a different structure than tree level - it appears 
as a rescattering effect only. It is IR finite and, in this model, is
proportional to the one loop result for $f_{1T}^{\perp}$~\cite{ji}
and ${\cal P}_y$ in BHS. The ratio of $h_{1T}(x,k_{T})$
to $h_{1T}^{\perp}(x,k_{T})$ will differ from the tree level.
When combined with a measure of transversely polarized quarks, the
fragmentation function $H_1^{\perp}(z)$, the
integrated $h_1(x)$ 
will contribute to the observable weighted meson azimuthal asymmetry 
from a transversely polarized nucleon~\cite{kotz0,boer}.

As mentioned in the introduction, the 
$T$-odd structure function $h_1^\perp(x,k_{T})$ is of great
interest theoretically,  since it vanishes at tree level, and 
experimentally, since its determination does not involve polarized
nucleons. Repeating the calculation above {\em without
nucleon polarization} leads to the result~\cite{gold_gam}
\small\be
\frac{\varepsilon_{+-\perp j}k_{T j}}{M}h_1^\perp(x,k_{T})&=&
\frac{e_1e_2g^2}{2(2\pi)^4}
\frac{(m+xM)(1-x)}{\Lambda(k^2_T)}
\varepsilon_{+-\perp j}k_{Tj}
\frac{1}{k_T^2}\ln\frac{\Lambda(k^2_T)}{\Lambda(0)}.
\ee\normalsize
It is  leading twist and IR finite. 
Being $T$-odd it will appear in SIDIS observables 
along with $T$-odd fragmentation functions.
In particular, the following weighted semi-inclusive 
DIS cross section projects out a leading
$\cos2\phi$ asymmetry~\cite{boer}, 
\small
\begin{equation}
{\langle {\vert P^2_{h{\perp}} \vert \over {M M_h}} \cos2\phi \rangle}_{UU}= 
\frac{\int d^2P_{h\perp} {\vert P^2_{h\perp}\vert \over {M M_h}}
\cos 2\phi d\sigma}
{\int d^2 P_{h\perp} d\sigma} = 
\frac{{8(1-y)} \sum_q e^2_q h^{\perp(1)}_1(x) z^2 H^{\perp(1)}(z)}
{{(1+{(1-y)}^2)}  \sum_q e^2_q f_1(x) D_1(z)}
\label{ASY} 
\end{equation}
\normalsize
where the subscript $UU$ indicates unpolarized beam and 
target(Note: The non-vanishing 
$\cos2\phi$ asymmetry originating from kinematical and 
dynamical effects only appears at order $1/Q^2$~\cite{CAHN,kotz0,OABD}).
The functions $h_1^{\perp (1)}(x)$ and $H_1^{\perp (1)}(z)$ are the weighted
moments of the distribution and fragmentation functions 
\small
\be
h^{\perp(1)}_1(x) \equiv \int d^2k_T \frac{k^2_T}{2M^2} h^{T}_1(x,k_T),\quad
H^{\perp(1)}_1(z) \equiv z^2 \int d^2p_T \frac{k^2_T}{2M^2_h} 
H^{T}_1(z,-p_T).
\ee\normalsize
\begin{figure}[t]
\includegraphics[height=6.0 cm]{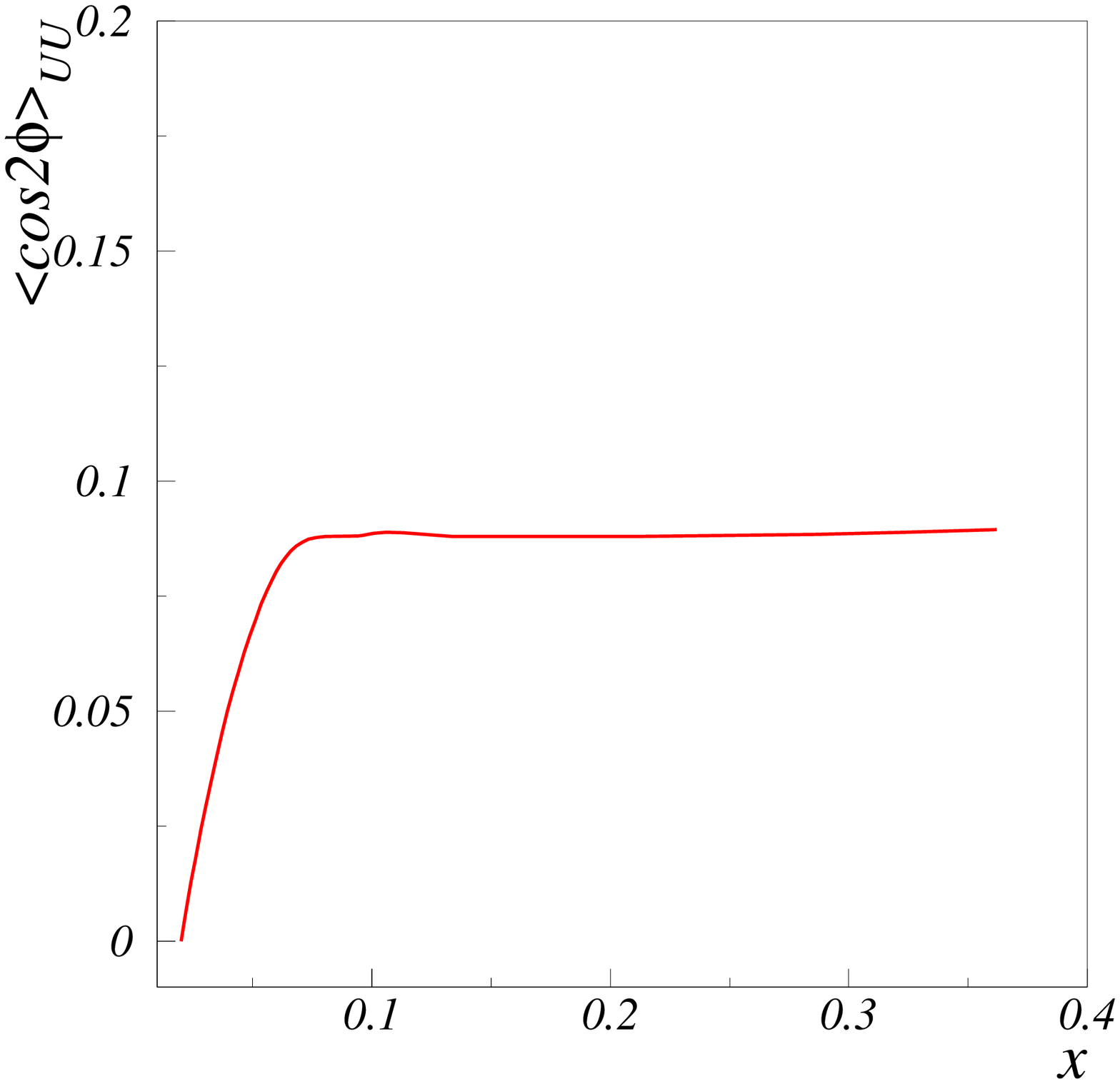} 
\includegraphics[height=6.0 cm]{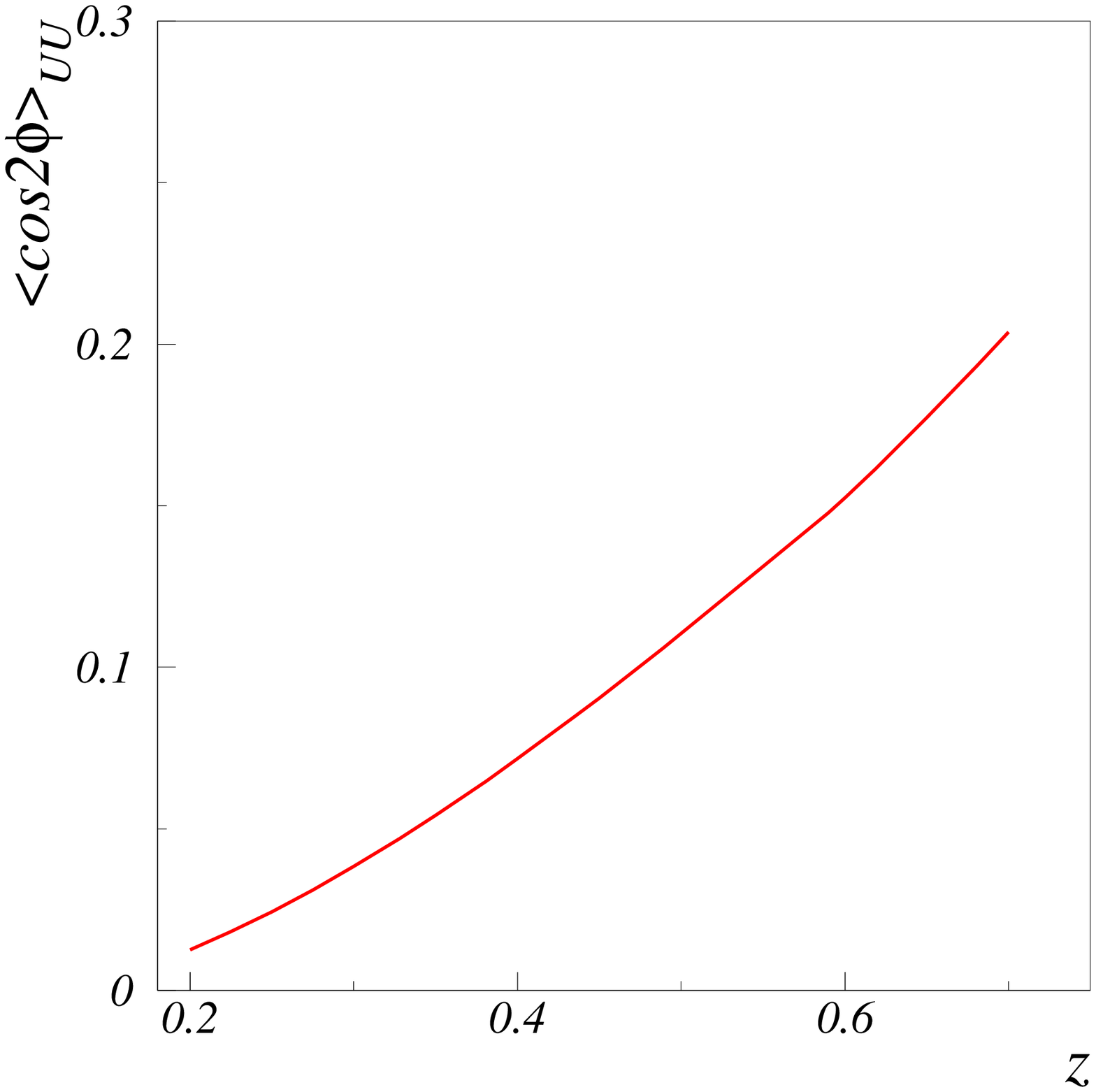}
\caption{Left Panel: The \protect{${\langle \cos2\phi \rangle}_{UU}$} 
asymmetry for 
\protect{$\pi^+$} production 
as a function of \protect{$x$}. Right Panel:
The \protect{${\langle \cos2\phi \rangle}_{UU}$} asymmetry for 
\protect{$\pi^+$} production 
as a function of \protect{$z$}.}
\label{A}
\end{figure}
$k_T$, $p_T$,  are the transverse momentum of the quark
in the target proton, and fragmenting quark respectively
and $M$, $M_h$ are the mass of the target proton and produced hadron.
We evaluate the ${\langle \cos2\phi \rangle}_{UU}$ asymmetry obtained 
from the  approximation,
\small\be
{\langle \cos2\phi \rangle}_{UU} \approx {{M M_h} \over 
{\langle P^2_{h\perp} \rangle} } {\langle {{\vert P^2_{h\perp} \vert} 
\over {M M_h}} \cos2\phi \rangle}_{UU} 
\label{EXP}
\ee
\normalsize
in the  HERMES kinematic range corresponding to 
$1$ GeV$^2$ $\leq Q^2 \leq 15$ GeV$^2$, $4.5$ GeV $\leq E_{\pi} \leq 13.5$ 
GeV, $0.2 \leq z \leq 0.7$, $0.2 \leq y \leq 0.8$, and taking $\langle 
P^2_{h\perp} \rangle = 0.25$ GeV$^2$ as input. 
The Collins ansatz~\cite{cnpb92,kotz0} for the analyzing power of 
transversely polarized quark fragmentation 
function $H^{\perp(1)}_1(z)$, has been  adopted~\cite{OBDN}. 
For $D_1(z)$, the simple parameterization from Ref.~\cite{REYA} was used. 
In Fig\ref{A} the ${\langle \cos2\phi \rangle}_{UU}$ of 
Eq.(\ref{ASY}) for $\pi^+$ production on a proton target is presented 
as a function of $x$ and $z$, respectively. 
Using $\Lambda_{QCD}=0.2\ GeV$ and $\mu=0.8\ GeV$,
Fig.\ref{A} indicates that the $\cos2\phi$ asymmetry related to  
is large enough (about $8\%$) to be measured~\cite{ggprep}.  
%%%%%%%%%%%%%%%%%%%%%%%%%%%%%%%%%%%%%%%%%%%%%%%%
%% BACKMATTER
%%%%%%%%%%%%%%%%%%%%%%%%%%%%%%%%%%%%%%%%%%%%%%%%
\subsection{Conclusions}
\vskip-.5cm
The interdependence of intrinsic transverse
quark momentum and angular momentum conservation are intimately
tied with the  studies of transversity. 
This is demonstrated from
analyses of the tensor charge in the context
of the vector dominance approach to exclusive meson photo-production,
to SSAs in SIDIS\cite{gamb_gold}.  
In the study of the tensor charge we find
 $\left< k_{T}^2 \right>$ factor that appears in
rescattering models in meson photoproduction
where interference phenomena are 
non-zero due to rescattering.
In the case of unpolarized beam and target we have predicted at HERMES
energies the sizable $\cos2\phi$ asymmetry associated with the
asymmetric distributions of transversely polarized quarks inside
of unpolarized hadrons.
{\small
\subsubsection{Acknowledgements}\vskip-.25cm
L.G. thanks the organizers of SPIN 2002 for the invitation to present
this work. I also thank 
Daniel Boer, Dennis Sivers and Feng Yuan for useful discussions.}
%%%%%%%%%%%%%%%%%%%%%%%%%%%%%%%%%%%%%%%%%%%%%%%%
%% You may have to change the BibTeX style below, depending on your
%% setup or preferences.
%%
%% If the bibliography is produced without BibTeX comment out the
%% following lines and see the aipguide.pdf for further information.
%%
%% For The AIP proceedings layouts use 
\bibliographystyle{aipproc}   % if natbib is available
%\bibliographystyle{aipprocl} % if natbib is missing

%%%%%%%%%%%%%%%%%%%%%%%%%%%%%%%%%%%%%%%%%%%
%% You probably want to use your own bibtex database here
%%%%%%%%%%%%%%%%%%%%%%%%%%%%%%%%%%%%%%%%%%%
%\vskip-0.25
\bb{99}
\bibitem{jaffe91}
X.\ Artu and M.\ Mekhfi, Z. Phys. C45 (1990) 669;
R.\ L.\ Jaffe and X.\ Ji, Phys. Rev. Lett. 67 (1991) 552;
Nucl. Phys. B375 (1992) 527.
\bibitem{ralston79}
J. Ralston and D. E. Soper, Nucl. Phys. B152 (1979) 109.
\bit{cnpb92} J.C.~Collins, Nucl. Phys. B396, 161 (1993).
\bibitem{kotz0} A. M. Kotzinian, Nucl. Phys B441 (1995) 234.
\bit{ANSL} M. Anselmino and F. Murgia, Phys. Lett B442 (1998) 470;
M. Anselmino, V. Barone, A. Drago and F. Murgia, hep-ph/0209073.
\bibitem{cplb} J. C. Collins, Phys. Lett. B536 (2002) 43.
\bibitem{mulders2} R. D. Tangerman and P. J. Mulders,
Phys. Lett. B352 (1995) 129; Phys. Rev. D51 (1995) 3357;
Nucl. Phys. B461 (1996) 197.
\bibitem{hermes} A. Airapetian {\it et al.}, Phys. Rev. Lett.
84 (2000) 4047; A. Bravar, Nucl. Phys. Proc. Suppl. 79
(1999) 520.
\bibitem{diehl} P. Hoodbhoy and X. Ji, Phys. Rev. D58 (1998) 054006;
M. Diehl, Eur. Phys. J. C19 (2001) 485.
\bibitem{gamb_gold} L. Gamberg and G. R. Goldstein, Phys.
Rev. Lett. 87 (2001) 242001.
\bibitem{gold_owens} G. R. Goldstein and J. F. Owens,
Phys. Rev. D7 (1973) 865; Nucl. Phys. B71 (1974) 461.
\bibitem{brodsky}
S. Brodsky, D.S. Hwang and I. Schmidt, Phys. Lett. B530 (2002) 99.
\bibitem{ji} X. Ji and F. Yuan, Phys. Lett. B543 (2002) 66; 
A.V. Belitsky, {\it et al.}, hep-ph/0208038.
\bit{sivers} D. Sivers, Phys. Rev D 41 (1990) 83 ; Phys. Rev. D 43 (1991) 261.
\bibitem{kotz} A. M. Kotzinian and P. J. Mulders,
Phys. Lett. B406 (1997) 373.
\bibitem{boer} D. Boer and P. J. Mulders, Phys. Rev. D57 (1998) 5780.
\bibitem{gold_gam} G. R. Goldstein and L. Gamberg,
hep-ph/0209085. 
\bit{CAHN} R.N. Cahn, Phys. Lett.  B78 (1978) 269; Phys. Rev. D40 (1989) 
3107. 
\bit{OABD} K.A.~Oganessyan, et.al., Eur. Phys. J.  C5 (1998) 681.
\bit{OBDN} K.A.~Oganessyan, N. Bianchi, E. De Sanctis, and 
W.D. Nowak, Nucl. Phys. A689 (2001) 784.
\bit{REYA} E.~Reya, Phys. Rep. 69 (1981) 195.  
\bibitem{ggprep} L. Gamberg, G. R. Goldstein and 
K.A.~Oganessyan,  in preparation.

\eb

%%%%%%%%%%%%%%%%%%%%%%%%%%%%%%%%%%%%%%%%%%%
%% Just a reminder that you may have to run bibtex
%% All of it up to \end{document} can be removed
%% if you don't like the warning.
%%%%%%%%%%%%%%%%%%%%%%%%%%%%%%%%%%%%%%%%%%%
\IfFileExists{\jobname.bbl}{}
 {\typeout{}
  \typeout{******************************************}
  \typeout{** Please run "bibtex \jobname" to optain}
  \typeout{** the bibliography and then re-run LaTeX}
  \typeout{** twice to fix the references!}
  \typeout{******************************************}
  \typeout{}
 }

\end{document}